\input harvmac
\def\half{{1 \over 2}}

\def\>{{\rangle}}
\def\<{{\langle}}

\def\p{{\partial}}

\def\s{{\sigma}}

\def\a {{\alpha}}
\def\b {{\beta}}

\def\ad {{\dot \a}}

\def\d {{\delta}}

\def\e {{\epsilon}}

\def\k {{\kappa}}

\def\t {{\theta}}

\def \t {{\theta}}

\def \tb {{\bar\theta}}

\Title{\vbox{\hbox{IFT-P.007/97}}}
{\vbox{\centerline{{\bf Conformal Compensators and Manifest
Type IIB S-Duality }}}}  
\bigskip\centerline{Nathan Berkovits}
\bigskip\centerline{Instituto
de F\'{\i}sica Te\'orica, Univ. Estadual Paulista}
\centerline{Rua Pamplona 145, S\~ao Paulo, SP 01405-900, BRASIL}
\bigskip\centerline{e-mail: nberkovi@ift.unesp.br}
\vskip .2in
Using the conformal compensator superfields of N=2 D=4 supergravity, 
the Type IIB S-duality transformations are expressed as a linear rotation 
which mixes the compensator and matter superfields. The classical superspace 
action for D=4 compactifications of Type IIB supergravity is manifestly 
invariant under this transformation. Furthermore, the introduction of 
conformal compensators allows a Fradkin-Tseytlin term to be added to the 
manifestly SL(2,Z)-covariant sigma model action of Townsend and Cederwall. 

\Date{January 1998}
\newsec {Introduction}

Although the evidence for S-duality of the Type IIB superstring is continually
growing, an explanation of this symmetry is still lacking. One cause of
the difficulty is that the D=10 Type IIB supergravity action is poorly
understood, both because of the chiral four-form and because of the lack
of an off-shell D=10 superspace formalism. Another cause is that S-duality
transformations take their simplest form in `Einstein gauge', whereas the
superstring is easiest to study in `string gauge'. 

To avoid these problems, Type IIB S-duality will be studied in this paper
for N=2 D=4 theories which are obtained by compactification of the 
Type IIB superstring on a Calabi-Yau manifold. Since off-shell
N=2 D=4 superspace is well understood, it is straightforward to 
construct superspace actions for these N=2 D=4 supergravity theories. 
The superspace actions involve conformal compensators which will permit a 
conformally gauge-invariant definition of S-duality transformations. 

The usual superspace procedure for coupling 
to supergravity is to first introduce
conformal compensators which allow the action in a flat metric
to be invariant under global conformal transformations. 
One then covariantly
couples to conformal supergravity
and finally, chooses a conformal-breaking condition which turns the
conformal supergravity multiplet into a Poincar\'e supergravity multiplet.
The choice of conformal-breaking
condition determines if one is in `Einstein gauge', `string gauge', or some
other gauge. 

In the second section of this paper, it will be shown that Type IIB 
S-duality transformations
take their simplest form before choosing a conformal-breaking condition, 
when they linearly rotate the conformal compensator hypermultiplet into the
`universal' hypermultiplet and leave all other multiplets unchanged. 
It is easy to prove that the classical superspace action obtained by
compactifying the Type IIB superstring on a Calabi-Yau manifold 
is invariant under this transformation. 

Recently, Townsend and Cederwall have proposed a manifestly SL(2,Z)-covariant
sigma model action for the Type IIB Green-Schwarz superstring
\ref\town{P.K. Townsend,
{\it Membrane Tension and Manifest IIB S-Duality}, hep-th/9705160\semi 
M. Cederwall and P.K. Townsend,
{\it The Manifestly SL(2,Z)-Covariant Superstring}, hep-th/9709002.}. 
Like the usual
Green-Schwarz sigma model, their action lacks a Fradkin-Tseytlin term
which couples the spacetime dilaton to the worldsheet curvature\ref\ft
{E. Fradkin and A. Tseytlin, {\it Effective Field Theory
from Quantized Strings}, Phys. Lett. B158 (1985) 316}. 
But previously, 
a sigma model action which includes a Fradkin-Tseytlin term was constructed  
using a modified Green-Schwarz description of the Type IIB
superstring compactified on a Calabi-Yau manifold \ref\ber{N. Berkovits,
{\it Covariant Quantization of the Green-Schwarz
Superstring in a Calabi-Yau Background}, Nucl. Phys. B431 (1994) 258,
hep-th/9404162.}\ref\bers{
N. Berkovits
and W. Siegel, {\it Superspace Effective Actions for 4D Compactifications
of Heterotic and Type II Superstrings}, 
Nucl. Phys. B462 (1996) 213, hep-th/9510106.}. 
In the third section of this paper, these two
actions will be combined to form a manifestly SL(2,Z)-covariant sigma model
action which includes a Fradkin-Tseytlin term. 

\newsec{ Conformal Compensators}

\subsec{Calabi-Yau compactififaction of the Type IIB superstring}

For compactifications of the Type IIB superstring on a six-dimensional
Calabi-Yau manifold with $h_{2,1}$ complex moduli and $h_{1,1}$
Kahler moduli, the massless N=2 D=4 supersymmetry multiplets include 
an N=2 D=4 supergravity multiplet, $h_{2,1}$ vector multiplets, and
$h_{1,1}+1$ hypermultiplets where the $+1$ comes from the `universal'
hypermultiplet. To construct covariant actions with manifest
spacetime supersymmetry, it is convenient to split the supergravity
multiplet into a conformal supergravity multiplet and conformal
compensator multiplets. 

If the action in a flat metric is invariant under global conformal
transformations, one makes the action super-reparameterization invariant
by covariantly coupling the action to conformal supergravity. 
If the action in a flat metric is not invariant under global conformal
transformations, one first couples to the conformal compensators in
such a way that the transformation of the compensators cancels the
transformation of the action. One then couples the combined
action to conformal supergravity. Gauge-fixing the conformal invariance 
turns the conformal supergravity multiplet into a Poincar\'e supergravity
multiplet, but complicates the supersymmetry transformations.

Although there is some ambiguity in the choice of conformal compensator
multiplets for N=2 D=4 supergravity \ref\dew
{B. de Wit, P.G. Lauwers and A. Van Proeyen, {\it
Lagrangians of N=2 Supergravity-Matter Systems},
Nucl. Phys. B255 (1985) 569.}, superstring field theory implies
that these compensator multiplets consist of a vector multiplet and 
a tensor hypermultiplet \bers\ref\sieg{
W. Siegel, {\it Curved Extended Superspace from Yang-Mills Theory
a la Strings}, Phys. Rev. D53 (1996) 3324, hep-th/9510150.}. 
Superstring field theory also implies that
the $h_{1,1}+1$ hypermultiplets are all 
tensor hypermultiplets (as opposed to scalar hypermultiplets).
Conveniently, actions involving tensor hypermultiplets 
are much
easier to construct in N=2 D=4 superspace than actions involving scalar
hypermultiplets. Note that component versions of
scalar hypermultiplet actions coming from Type IIB compactifications have
been extensively studied in various important papers which include \ref\cec
{S. Cecotti, S. Ferrara and L. Girardello,
{\it Geometry of Type II Superstrings and the Moduli of Superconformal
Field Theories}, Int. Journ. of Mod. Phys. A4 (1989) 2475.}, 
\ref\sad {S. Ferrara and S. Sabharwal,
{\it Quaternionic Manifolds for Type II Superstring Vacua of
Calabi-Yau Spaces}, Nucl. Phys. B332 (1990) 317.}, and \ref\van
{B. de Wit and A. Van Proeyen, {\it Special Geometry, Cubic Polynomials
and Homogeneous Quaternionic Spaces}, Commun. Math. Phys. 149 (1992) 307.}.

\subsec{N=2 D=4 superspace}

The variables of N=2 D=4 superspace are [$x^\mu$, $\t^\a_j$, $\tb^{j \ad}$]
where $\mu=0$ to 3, $\a$ and $\ad$ = 1 to 2, and $j=1$ to 2 is
an internal $SU(2)_R$ index which is raised and lowered using the
anti-symmetric $\e^{jk}$ tensor. $\tb^{j \ad }$ is the complex conjugate
of $\t^\a_j$ and under $U(1)_R$ transformations,
$\t^\a_j$ carries $+1$ charge and $\tb^{j\ad}$ carries $-1$ charge.
Under global conformal transformations, $\t^\a_j$ and $\tb^{j\ad}$ carry 
scale-weight $-\half$ and $x^\mu$ carries scale-weight $-1$. 
In a flat metric, supersymmetric derivatives are defined as
\eqn\aaa{D_\a^j
= {\p \over {\p \t^\a_j}} + i \tb^{j\ad } \s^\mu_{\a\ad} \p_\mu,\quad
\bar D^\ad_j =
{\p \over {\p \tb_\ad^j}} + i \t_{j\a} \bar\s_\mu^{\ad\a} \p^\mu.}
There are three types of N=2 D=4 multiplets which will be useful to review:
the vector multiplet, the tensor hypermultiplet, and the conformal
supergravity multiplet. 

The field-strength of a vector multiplet is described by a restricted
chiral
superfield $W$ satisfying
\eqn\chir{\bar D_\ad^+ W =\bar D_\ad^-W =  0,}
$$ D_\a^{(j} D^{k) \a} W =
\bar D_\ad^{(j} D^{k) \ad} \bar W $$
where the first constraint implies that $W$ is chiral/chiral,
while the second constraint
implies that $W$ is restricted.
The physical bosonic components of $W$ appear as
\eqn\physw{
W = w(x) + \t_j^{(\a} \t^{\b) j} \s^{\mu\nu}_{\a\b} F_{\mu\nu} (x) + ... }
where $w$ is a complex scalar and $F_{\mu\nu}$ is the vector field strength.
Under $U(1)_R\times SU(2)_R$, $W$ transforms as $(+2,1)$, so
$w$ and $\bar w$ transform as $(+2,1)$ and $(-2,1)$ while $F_{\mu\nu}$
transforms as $(0,1)$. Under conformal transformations,
$w$ has scale-weight $+1$ and $F_{\mu\nu}$ has scale-weight
$+2$. 

The field-strength of a tensor hypermultiplet is described by
a linear superfield $L_{jk}$ symmetric in its SU(2) indices which
satisfies the reality condition $L_{jk}=(L^{jk})^*$ and the linear constraint
\eqn\aab{ D^\a_{(j} L_{kl)} = 0, \quad
\bar D^\ad_{(j} L_{kl)} = 0.}
The physical bosonic components of $L_{jk}$ appear as
\eqn\physl{L_{jk} = l_{jk}(x) + \t_{(j}^\a \tb_{k)}^\ad
\e_{\mu\nu\rho\kappa}\s^\mu_{\a\ad} H^{\nu\rho\kappa}(x) +
... }
where $l_{jk}$ is a triplet of scalars transforming as
$(0,3)$ under $U(1)_R\times SU(2)_R$ and $H^{\mu\nu\rho}$ is the tensor
field-strength which transforms as $(0,1)$.
Under conformal transformations, $l_{jk}$ has scale-weight $+2$ and
$H_{\mu\nu\rho}$ has scale-weight $+3$. 

Although the constraints of \aab\ appear very different from the
constraints of \chir, they are actually closely related. This can be seen
by noting that the constraints of \aab\ imply that $L_{++}$ is restricted
twisted-chiral since it satisfies
\eqn\twist{D_+^\a L_{++} =\bar D^{\ad}_+ L_{++}=0,}
$$D_-^\a D_{-\a} L_{++} = D_+^\a D_{+\a}
L_{--},\quad
\bar D_-^\ad \bar D_{-\ad} L_{++} = \bar D_+^\ad \bar D_{+\ad} L_{--}.$$
The first two constraints imply that $L_{++}$ is chiral/anti-chiral, while
the second two constraints imply that $L_{++}$ is restricted.

Finally, the conformal supergravity multiplet is described by a supervierbein
superfield $E_A^M$ where $A$ denotes tangent-space vector and spinor indices
while $M$ denotes curved-space vector and spinor indices. The superfield
$E_A^M$ is subject to various torsion constraints which will not be
directly relevant for this paper. 

\subsec{S-duality in superspace}

The F-theory conjecture states that the Type IIB superstring compactified
on $\cal M$ is equivalent to F-theory compactified on $T_2 \times
{\cal M}$ with the complex modulus of $T_2$ parameterized by
$\tau = a -i e^{-\phi}$ where $a$ is the axion and $e^{-\phi}$ is the
dilaton. So choosing $\cal M$ to be the Calabi-Yau manifold, modular
invarince of $T_2$ implies that the compactified theory is invariant
under the S-duality SL(2,Z) transformation 
\eqn\sdual{\tau \to {{A \tau +B}\over{C\tau +D}}}
where A,B,C,D are integers satisfying $AD-BC=1$.

A natural question is how do the supersymmetry multiplets transform under
\sdual. 
For compactification on a Calabi-Yau manifold, 
the massless N=2 D=4 superfields include 
a compensating vector multiplet
described by $W^{(0)}$, 
physical vector multiplets described
by $W^{(X)}$ where $X=1$ to $h_{2,1}$, 
a compensating tensor hypermultiplet
described by $L_{jk}^{(0)}$, 
physical tensor hypermultiplets described by
$L_{jk}^{(Y)}$ for $Y$=1 to $h_{1,1}$, a physical `universal' hypermultiplet
described by $L'_{jk}$, and the conformal supergravity multiplet described
by $E_A^M$.  
As will be
shown below, the S-duality transformations of \sdual\ transform these
superfields as
\eqn\ssuper{ E_A^M \to E_A^M, 
\quad W^{(0)} \to W^{(0)},
\quad W^{(X)} \to W^{(X)},
\quad L_{jk}^{(Y)} \to L_{jk}^{(Y)},}
$$L^{(0)}_{jk} \to A L^{(0)}_{jk} + B L'_{jk}, \quad 
L'_{jk} \to C L^{(0)}_{jk} + D L'_{jk}.$$

To verify \ssuper, one first needs to determine 
how the components of the various superfields depend on $a$ and $\phi$.
Since the N=2 D=4 superconformal group includes local
$SU(2)_R\times U(1)_R$ rotations, 
one can gauge to zero the component fields $Im (w^{(0)})$, 
$Im (l_{jk}^{(0)})$, $l'_{++}$ and $l'_{--}$. This still leaves
local conformal transformations which can be used to gauge-fix
$l'_{+-}=1$.

In this gauge, the equation of motion for the scalar in the conformal
supergravity multiplet implies that $Re (w^{(0)})$ is 
on-shell a function of the other fields. So besides the scalars
coming from $W^{(X)}$ and $L^{(Y)}_{jk}$, there are four independent
scalars which can be defined in terms of $l^{(0)}_{+-}$, $Re(l^{(0)}_{++})$,
and the duals of $H^{(0)}_{\mu\nu\rho}$ and  
$H'_{\mu\nu\rho}$. In terms of the original bosonic fields of
D=10 Type IIB supergravity, 
\eqn\comp{ Re (l_{++}^{(0)}) = e^{-\phi}, \quad
l_{+-}^{(0)} = a, \quad
H_{\mu\nu\rho}^{(0)} = \p_{[\mu} \tilde b_{\nu\rho]}, }
$$l_{++}^{(Y)} = e^{-\phi} (G^{(Y)J\bar K} g_{J\bar K} + i
B^{(Y)J\bar K} b_{J\bar K}), \quad
l_{+-}^{(Y)} = 
B^{(Y)J\bar K} (\tilde b_{J\bar K}
- a~ b_{J\bar K}), $$
$$H_{\mu\nu\rho}^{(Y)} = 
B^{(Y)J\bar K}\p_{[\mu} A_{\nu\rho] J \bar K}, \quad 
H'_{\mu\nu\rho} = 
\p_{[\mu} b_{\nu\rho]},$$
$$F_{\mu\nu}^{(0)} = 
\omega^{J K L}\p_{[\mu} A_{\nu] J K L} + c.c., $$
$$w^{(X)}= h^{(X)JK} g_{JK}, \quad 
F_{\mu\nu}^{(X)} = 
h_{\bar K}^{(X)J} 
\bar\omega^{\bar K\bar L\bar M}\p_{[\mu} A_{\nu] J\bar L\bar M} + c.c., $$
where $J,K$ and $\bar J,\bar K$ are the complex coordinates of
the Calabi-Yau manifold ($J,K=1$ to 3), $G^{(Y)J\bar K}$ and     
$B^{(Y)J\bar K}$ are the Kahler moduli and torsion of the Calabi-Yau
manifold, 
$h^{(X)J K}$ are the complex moduli of the Calabi-Yau
manifold, $a$ and $e^{-\phi}$ are the D=10 axion and dilaton,
$g_{mn}$ is the D=10 metric ($m,n$ can point either 
in the spacetime directions or in the Calabi-Yau directions), 
$b_{mn}$ and $\tilde b_{mn}$ are the D=10 NS-NS and R-R two-forms, and 
$A_{mnpq}$ is the D=10 self-dual four-form. Note that 
$l_{++}^{(Y)}$ has $e^{-\phi}$ dependence since the Kahler moduli are
give by the conformally-invariant quantities 
$l_{++}^{(Y)}/
l_{++}^{(0)}$. 

To prove that \ssuper\ correctly defines the S-duality transformation,
first consider the shift transformation when $A=B=D=1$ and
$C=0$. Under this transformation, 
$L^{(0)}_{jk} \to  L^{(0)}_{jk} +  L'_{jk}$ and all other superfields
remain unchanged. Comparing with \comp, this implies that
\eqn\shift{ a\to a +1,\quad \tilde b_{mn} \to \tilde b_{mn} + b_{mn},}
which is the desired transformation. 

Now consider the strong/weak transformation when $A=D=0$, $B=-1$ and 
$C=1$. Under this transformation, 
$L^{(0)}_{jk} \to  -L'_{jk}$ and 
$L'_{jk} \to  L^{(0)}_{jk}$, which does not preserve the gauge-fixing
condition $l'_{jk}=\delta_{jk}$ (i.e. $l'_{++}=l'_{--}=0$, $l'_{+-}=1$).

So to obtain the transformations of the component fields, one needs
to perform a local $SU(2)_R$ and conformal transformation to return to
the original gauge choice. Alternatively, one can express the component
fields in a form which is invariant under $SU(2)_R$ and conformal
transformations, e.g. 
\eqn\invc{a = {{l^{(0)}\cdot l'}\over {l'\cdot l'}},\quad 
e^{-2\phi} + a^2  = {{l^{(0)}\cdot l^{(0)}}\over {l'\cdot l'}}}
where $A\cdot B\equiv A_{jk} B^{jk}$. Under the strong/weak
transformation,
\eqn\swt{
a \to {{-l'\cdot l^{(0)}}\over {l^{(0)}\cdot l^{(0)}}} =
{{-a}\over 
{e^{-2\phi} + a^2}},}
$$e^{-2\phi} + a^2  \to {{l'\cdot l'}\over {l^{(0)}\cdot l^{(0)}}}
= {1\over {e^{-2\phi} + a^2}},$$
which implies that $a+ i e^{-\phi} \to
-(a+ i e^{-\phi} )^{-1}$ as desired. 

Similarly, one can show that the other component fields transform
appropriately, e.g. 
\eqn\other{g_{mn} \to 
\sqrt{e^{-2\phi} + a^2} ~g_{mn},\quad 
b_{mn} \to -\tilde b_{mn},\quad 
\tilde b_{mn} \to b_{mn},\quad 
A_{mnpq} \to A_{mnpq}.}
Note that there are no 
$(e^{-2\phi} + a^2)$
factors in the transformations of $b_{mn}$ and $A_{mnpq}$ 
since, in a curved background, the component fields appearing
in $L_{jk}$ and $W$ carry tangent-space indices and the transformation
of the vierbein absorbs the 
$(e^{-2\phi} + a^2)$
factors coming from the conformal rescaling. 

Since any S-duality transformation can be described by a product of
shift and strong/weak transformations, the transformation of \ssuper\
correctly reproduces \sdual. It will now be shown that the classical
superspace action for the Type IIB compactification is invariant under
\ssuper. 

\subsec{N=2 D=4 superspace actions}

Two-derivative actions for the vector multiplets and tensor
hypermultiplets can be written in manifestly
supersymmetric notation as \ref\karl{A. Karlhede, U. Lindstrom
and M. Ro\v{c}ek, {\it Self-Interacting Tensor
Multiplets in N=2 Superspace}, Phys. Lett. B147 (1984) 297\semi
W. Siegel, {\it Chiral Actions for N=2 Supersymmetric Tensor
Multiplets}, Phys. Lett. B153 (1985) 51.}
\eqn\act{
\int d^4 x|_{\t_j^\a=\tb_j^\ad=0} [(D_+)^2 (D_-)^2 f_V(W^{(I)}) ~+ ~
\oint_{\cal C} {{d\zeta}\over {2\pi i}}
(D_-)^2 (\bar D_-)^2 f_T(\tilde L^{(J)},\tilde L') ~+~c.c.]}
where $I=0$ to $h_{2,1}$, $J=0$ to $h_{1,1}$, 
$f_V$ and $f_T$ are arbitrary functions,
$\oint_{\cal C} {{d\zeta}\over {2\pi i}}$ is some contour integration,
and
\eqn\ttt{\tilde
L =  L_{++} +\zeta L_{+-} + \zeta^2 L_{--}.}
The hypermultiplet
contribution
to \act\ is supersymmetric where
\eqn\uuu{\d_Q f_T = [\xi_j^\a D_\a^j +
\bar\xi^j_\ad \bar D^\ad_j -2i(\xi\s^\mu\tb
+\bar\xi\bar\s^\mu\t)\p_\mu] f_T,}
since
$D_+^\a f_T = \zeta D_-^\a f_T$ and
$\bar D_+^\ad f_T = \zeta \bar D_-^\ad f_T$, so
$(D_-)^2 (\bar D_-)^2 \d_Q f_T$ is a total derivative in $x^\mu$. 

These actions are invariant under global $SU(2)_R\times U(1)_R$ and
conformal transformations when $f_V$ is of degree 2 
and $f_T$ is of degree 1 (i.e.  
$f_V (\lambda W^{(I)}) = \lambda^2 f_V(W^{(I)})$ and 
$f_T (\lambda \tilde L^{(J)}, \lambda\tilde L') 
= \lambda f_V(\tilde L^{(J)},\tilde L')$).
Although $SU(2)_R$ invariance is not manifest, it can be made manifest
by writing the hypermultiplet action as \karl
\eqn\manif{
\int d^4 x|_{\t_j^\a=\tb_j^\ad=0}
\oint_{\cal C} {{\epsilon_{jk} }\over {2\pi i}}\zeta^j d\zeta^k 
{{(v \cdot D)^2 (v\cdot \bar D)^2}\over {(\zeta\cdot v)^4}} 
f_T (\widehat L^{(J)}, \widehat L') + c.c. }
where $\widehat L = \zeta^j \zeta^k L_{jk}$ and 
$v^j$ is a fixed two-component constant. Note that \manif\ is
invariant under
\eqn\minv{ v^j \to c_1 v^j + c_2 \zeta^j, \quad \zeta^j \to c_3 \zeta^j}
for arbitrary constants $c_1, c_2$ and $c_3$, so one can choose
$v^+=0$ and $v^- = \zeta^+ =1.$ With this gauge choice, \manif\ becomes
the hypermultiplet action of \act. 

For generic choices of $f_T$, the action of \act\ is not invariant
under \ssuper. However, it will now be argued that for 
classical actions coming from Type IIB 
compactifications, $f_T$ takes the form  
\eqn\alw{f_T (\tilde L^{(Y)}, \tilde L')=
{{i~d_{Y_1 Y_2 Y_3} \tilde L^{(Y_1)} \tilde L^{(Y_2)} \tilde L^{(Y_3)}}\over
{\tilde L^{(0)} \tilde L'}}}
where $d_{Y_1 Y_2 Y_3}$ are real symmetric constants and the contour
$\cal C$ goes clockwise around the two values of $\zeta$ for which
$\tilde L'=0$. It will then be shown that this choice of $f_T$
produces action invariant under \ssuper. 

\subsec{Type IIB hypermultiplet action}

The M-theory conjecture states that the Type IIA superstring compactified
on a manifold $\cal M$ is equivalent to M-theory compactified on
$S_1\times\cal M$. So when $\cal M$ is the Calabi-Yau manifold, validity
of this conjecture implies that the Type IIA action can be obtained
from dimensional reduction on a circle of a D=5 action.
In such an action, one of the scalars in the N=2 D=4 vector
multiplets comes from the fifth component of a D=5
vector. Therefore, gauge invariance of the D=5 action implies that
the zero mode of these scalars decouples in the D=4 action. 

As shown in \cec, 
this implies that $f_V$ in the 
action for the Type IIA compactification must have the form
\eqn\vectwoa{f_V( W^{(0)}, W^{(Y)}) =
{{i~d_{Y_1 Y_2 Y_3} W^{(Y_1)} W^{(Y_2)} W^{(Y_3)}}\over
{ W^{(0)}}}}
where $d_{Y_1 Y_2 Y_3}$
is a real symmetric constant
and $Y=1$ to $h_{1,1}$. Note that the zero modes of the relevant scalars
are given by $Re (w^{(Y)}/w^{(0)})$. These zero modes decouple since, 
under $\d W^{(Y)} = c^{(Y)} W^{(0)}$ for real constants $c^{(Y)}$, 
the action changes by
\eqn\chang{-2 \int d^4 x|_{\t_j^\a=\tb_j^\ad=0}  Im [   (D_+)^2 (D_-)^2
d_{Y_1 Y_2 Y_3} c^{(Y_1)} W^{(Y_2)} W^{(Y_3)}]} 
which is a surface term. 

In references \cec\ and \sad, 
a relation was found 
connecting the perturbative
effective action for Type IIA and Type IIB compactifications
on the same Calabi-Yau manifold. This relation was later
extended to superspace in \bers\ and 
states that in the string
gauge $L'_{jk}=\d_{jk}$, the perturbative
Type IIA action is obtained from the
Type IIB action by replacing the superfields $W^{(I)}$ with 
$L_{++}^{(I)}$, $\bar W^{(I)}$ with $L_{--}^{(I)}$,
$L_{++}^{(J)}$ with
$W^{(J)}$, $L_{--}^{(J)}$ with
$\bar W^{(J)}$, and by swapping $\theta^+$ with $\bar\theta^-$ and 
$D_+$ with $\bar D_-$.\foot{ This symmetry relation is broken
non-perturbatively and was mistakenly called 
mirror symmetry in reference \bers. Mirror symmetry is believed to
be preserved non-perturbatively and relates Type IIB compactification
on a Calabi-Yau manifold with Type IIA compactification on the mirror
Calabi-Yau manifold.}

So in the gauge $L'_{jk}=\d_{jk}$, the hypermultiplet action for
the Type IIB compactification must have the form 
\eqn\must{ -2 ~Im[\int d^4 x|_{\t_j^\a=\tb_j^\ad=0} 
(D_-)^2 (\bar D_-)^2 
{{d_{Y_1 Y_2 Y_3} L_{++}^{(Y_1)} L_{++}^{(Y_2)} L_{++}^{(Y_3)}}\over
{ L_{++}^{(0)}}} ]. } It will now be shown that this comes
from gauge-fixing a hypermultiplet
action with $f_T$ and $\cal C$ defined as in \alw. 

In the string gauge $\tilde L' =\zeta$ (i.e. $L'_{jk}=\d_{jk}$), 
the contour $\cal C$ should
go clockwise around the origin and counter-clockwise at $\infty$. 
Note that the zero of $\tilde L'$ at $\zeta=\infty$ can be 
understood by taking the limit as $c\to 0$ of $\tilde L'=
c(\zeta -c) (\zeta + {1\over c})$. So the hypermultiplet action
defined with $f_T$ and $\cal C$ as in \alw\ is
\eqn\hype{
\int d^4 x|_{\t_j^\a=\tb_j^\ad=0} 
(\oint_{0}-\oint_{\infty}) {{d\zeta}\over {2\pi i}}
(D_-)^2 (\bar D_-)^2 
{{i~d_{Y_1 Y_2 Y_3} \tilde L^{(Y_1)} \tilde L^{(Y_2)} \tilde L^{(Y_3)}}\over
{\tilde L^{(0)} \zeta}}}
$$= 
\int d^4 x|_{\t_j^\a=\tb_j^\ad=0} 
[(D_-)^2 (\bar D_-)^2 
{{i~d_{Y_1 Y_2 Y_3}L_{++}^{(Y_1)} L_{++}^{(Y_2)} L_{++}^{(Y_3)}}
\over
{L_{++}^{(0)} }}$$
$$-
\oint_{\infty} {{d\zeta}\over{2\pi i\zeta}}
({D_+\over \zeta})^2 ({\bar D_+\over\zeta})^2 
{{i~d_{Y_1 Y_2 Y_3} \tilde L^{(Y_1)} \tilde L^{(Y_2)} \tilde L^{(Y_3)}}\over
{\tilde L^{(0)}}} ]$$
$$= 
\int d^4 x|_{\t_j^\a=\tb_j^\ad=0} 
[(D_-)^2 (\bar D_-)^2 
{{i~d_{Y_1 Y_2 Y_3}L_{++}^{(Y_1)} L_{++}^{(Y_2)} L_{++}^{(Y_3)}}
\over
{L_{++}^{(0)} }}$$
$$-(D_+)^2 (\bar D_+)^2 
{{i~d_{Y_1 Y_2 Y_3}L_{--}^{(Y_1)} L_{--}^{(Y_2)} L_{--}^{(Y_3)}}
\over
{L_{--}^{(0)} }}],$$
which agrees with \must. 

Note that when the Calabi-Yau manifold is the mirror of another
Calabi-Yau manifold, the effective action can be obtained from either a
Type IIA or Type IIB compactifications, so it must be of the form
\eqn\mirror{
\int d^4 x|_{\t_j^\a=\tb_j^\ad=0} 
 [ (D_+)^2 (D_-)^2
{{i~\tilde d_{X_1 X_2 X_3} W^{(X_1)} W^{(X_2)} W^{(X_3)}}\over{W^{(0)}}} }
$$
+~\oint_{\cal C} {{d\zeta}\over {2\pi i}}
(D_-)^2 (\bar D_-)^2 
{{i~d_{Y_1 Y_2 Y_3} \tilde L^{(Y_1)} 
\tilde L^{(Y_2)} \tilde L^{(Y_3)}}\over
{\tilde L^{(0)} \tilde L'}} + c.c. ]$$
where $\tilde d_{X_1 X_2 X_3}$ are the symmetric constants on
the mirror manifold and $\cal C$ goes clockwise
around the two zeros of $\tilde L'$.

Finally, it will be shown that the Type IIB 
action is invariant under the
S-duality transformations of \ssuper\ when $f_T$ and $\cal C$ are
defined as in \alw. 

Under the shift transformation 
$L^{(0)}_{jk} \to  L^{(0)}_{jk} +  L'_{jk}$, 
\eqn\sinv{f_T \to f_T + 
{{d_{Y_1 Y_2 Y_3} \tilde L^{(Y_1)} 
\tilde L^{(Y_2)} \tilde L^{(Y_3)}}\over
{(\tilde L^{(0)})^2}} ~( -1 - {{\tilde L'}\over \tilde L^{(0)}} - 
({{\tilde L'}\over \tilde L^{(0)}})^2 - ...)}
and since $\d f_T$ has no poles when $\tilde L'=0$, the contour integral
$\oint_{\cal C} d\zeta $
$(D_-)^2 (\bar D_-)^2 ~\d f_T $ vanishes. 

Under the strong/weak transformation
defined by 
$L^{(0)}_{jk} \to  
-L'_{jk}$ and 
$L'_{jk} \to 
L^{(0)}_{jk}$, $f_T \to -f_T$ where the contour $\cal C$ now goes
clockwise around the two values of $\zeta$ for which $\tilde L^{(0)}=0$.   
But since the only poles of $f_T$ occur when $\tilde L^{(0)}=0$ or when
$\tilde L'=0$, one can deform the contour off the zeros of $\tilde L^{(0)}$
until they go counter-clockwise around the zeros of $\tilde L'$.
Finally, reversing the direction of the contour cancels the minus
sign to give the original expression. 

\newsec{S-Dual Fradkin-Tseytlin Term}

Just as compactification to D=4 simplifies the analysis of S-duality
transformations, it also simplifies quantization of the superstring. 
For Calabi-Yau compactifications of the Type II superstring to D=4,
the four fermionic $\kappa$-symmetries can be interpreted
as N=(2,2) worldsheet superconformal invariances \ref\twist
{D.P. Sorokin, V.I. Tkach, D.V. Volkov and A.A. Zheltukhin,
{\it From the Superparticle Siegel Symmetry to the Spinning Particle
Proper Time Supersymmetry}, Phys. Lett. B216 (1989) 302\semi
N. Berkovits, {\it A Covariant Action for the Heterotic
Superstring with Manifest Space-Time Supersymmetry and Worldsheet
Superconformal Invariance}, Phys. Lett. B232 (1989) 184\semi
M. Tonin, {\it Worldsheet Supersymmetric Formulations of Green-Schwarz
Superstrings}, Phys. Lett. B266 (1991) 312\semi
E.A. Ivanov and A.A. Kapustnikov, {\it Towards a Tensor Calculus for
Kappa Supersymmetry}, Phys. Lett. B267 (1991) 175\semi
N. Berkovits, {\it The Heterotic Green-Schwarz Superstring on an
N=(2,0) Super-Worldsheet}, Nucl. Phys. B379 (1992) 96.}. 
After slightly 
modifying the usual Green-Schwarz worldsheet variables,
the superstring can be quantized in worldsheet
conformal gauge with manifest N=2 D=4
super-Poincar\'e invariance \ber. 
Unlike the standard Green-Scwharz sigma model, the sigma model action
for this modified Green-Schwarz
superstring includes a Fradkin-Tseytlin term which couples
the spacetime dilaton to the worldsheet supercurvature \ref\hets
{N. Berkovits, {\it A New Sigma Model Action for the Four-Dimensional
Green-Schwarz Heterotic Superstring}, Phys. Lett. B304 (1993) 249.}\bers. 

For compactifications of 
the Type IIB superstring, this sigma model action in worldsheet
conformal gauge is given by \bers
\eqn\sigmam{{1\over{\alpha'}} \int dz d\bar z [ 
\eta_{ab} E_M^a E_N^b \p_z Y^M \p_{\bar z} Y^N + b'_{MN} \p_z 
 Y^M \p_{\bar z} Y^N  + ... ]  }
$$+~\int dz d\bar z  [d\kappa_+ d\bar\kappa_+ \Sigma_c \log (W) +
d\kappa_+ d\bar\kappa_- \Sigma_{tc} \log (L_{++}^{(0)})~ + ~c.c. ]$$
where $Y^M= [x^m,\theta_j^\a,\bar\theta_j^\ad]$, 
$b'_{MN}$ is the two-form potential whose field-strength is $L'_{jk}$
of the previous section, $\k_{\pm}$ and $\bar\k_{\pm}$ are the
anti-commuting variables of D=2 N=(2,2) superspace, 
$\Sigma_c$ and $\Sigma_{tc}$ are the chiral and twisted-chiral 
N=(2,2) worldsheet superfields whose top component is the worldsheet
curvature, and ... refers to terms (written explicitly in \bers) which
will not be relevant to the discussion. 

As argued in \bers, the first line in \sigmam\ is invariant under classical
N=(2,2) worldsheet superconformal invariance if $E_A^M$ satisfies
the torsion constraints of N=2 D=4 supergravity and the field-strength
for $b'_{MN}$ satisfies $L'_{jk}=\delta_{jk}$. The second line 
of \sigmam\ is not invariant under classical
worldsheet superconformal transformations, and as usual for
a Fradkin-Tseytlin term, its classical variation is
expected to cancel the quantum variation of the first line when the
background superfields are on-shell. 
This has been explicitly checked for the heterotic version of \sigmam\
in reference \ref\deb{J. de Boer and K. Skenderis, {\it Covariant
Computation of the Low-Energy Effective Action of the Heterotic
Superstring}, Nucl. Phys. B481 (1996) 129, hep-th/9608078.}. 

Recently, Townsend and Cederwall \town\ have constructed a manifestly 
SL(2,Z)-covariant action for the superstring by introducing two
worldsheet $U(1)$ gauge fields, ${\cal A}_i$ and $\tilde {\cal A}_i$.
For the $(p,q)$ superstring, the constants $p$ and $q$ are replaced
by worldsheet fields, $S$ and $\tilde S$, which
are the conjugate momenta to these worldsheet gauge fields. 

Like the standard Green-Schwarz sigma model action, their sigma model
lacks a Fradkin-Tseytlin term. But using the methods of \town\
and the results of the previous
section,
it is straightforward to generalize \sigmam\ to
an SL(2,Z)-covariant action in worldsheet conformal gauge. 
The appropriate generalization in conformal gauge is 
\eqn\signew{{1\over{\alpha'}} \int dz d\bar z [ ~
S F + \tilde S \tilde F +
\eta_{ab} E_M^a E_N^b \p_z Y^M \p_{\bar z} Y^N + ( S~b'_{MN} +
\tilde S ~ b_{MN}^{(0)}) \p_z 
 Y^M \p_{\bar z} Y^N  + ... ] }
$$+\int dz d\bar z  [d\kappa_+ d\bar\kappa_+ \Sigma_c \log (W) +
d\kappa_+ d\bar\kappa_- \Sigma_{tc} \log ( L_{++}^{(0)} L'_{+-}
- L_{+-}^{(0)} L'_{++}) ~+ ~c.c. ]$$
where $F$ and $\tilde F$ are the field-strengths for
${\cal A}_i$ and $\tilde {\cal A}_i$, $b^{(0)}_{MN}$ 
is the potential whose field-strength
is $L^{(0)}_{jk}$, and $L_{jk}^{(0)}$ is replaced with
$(L_{jk}^{(0)} L'_{+-}
- L_{+-}^{(0)} L'_{jk})$ everywhere it appears 
in $...$ .
It is easy to check that \signew\ is invariant
under the SL(2,Z) transformations
\eqn\chek{ S\to A S - B \tilde S,\quad \tilde S\to - CS + D\tilde S, \quad  
{\cal A}_i\to C {\cal A}_i + D \tilde {\cal A}_i,\quad 
\tilde {\cal A}_i\to A {\cal A}_i + B\tilde {\cal A}_i, }
$$ L^{(0)}_{jk}\to A 
L^{(0)}_{jk} + B L'_{jk},\quad 
L'_{jk}\to C 
L^{(0)}_{jk} + D L'_{jk},$$
where $AD-BC=1$. 

The equations of motion for ${\cal A}_i$ and $\tilde {\cal A}_i$ imply that 
$S$ and $\tilde S$ are constants on-shell, and since the gauge fields
are $U(1)$, these constants must be integer-valued and 
can be identified with $p$ and $q$ \ref\wit
{E. Witten, {\it Bound States of Strings and $p$-Branes}, Nucl. Phys.
B460 (1996) 335, hep-th/9510135.}. The transformations of the
spacetime superfields in \chek\ are the same as in \ssuper, so
the action of \signew\ correctly describes the $(p,q)$ superstring.

When $S$ and $\tilde S$ take background values $p$ and $q$, 
classical 
worldsheet superconformal invariance of the first line in \signew\ 
implies
that $p L'_{jk} + q L^{(0)}_{jk} = \delta_{jk}$. 
So to compare with the usual $(p,q)$ sigma model action, one needs to perform
a local conformal and $SU(2)_R$ transformation on all
background superfields in order to recover the string gauge
$L'_{jk}=\d_{jk}$. For example, when written
in terms of the string-gauge metric 
$\hat g_{\mu\nu}$, 
\eqn\comparing{ g_{\mu\nu} \p_z x^\mu \p_{\bar z} x^\nu =
\sqrt {(q e^{-\phi})^2 + (p + q a)^2}~ \hat g_{\mu\nu}
\p_z x^\mu \p_{\bar z} x^\nu ,}
reproducing the $(p,q)$ tension formula of \ref\shwa
{J.H. Schwarz, {\it An SL(2;Z) Multiplet of Type IIB Superstrings},
Phys. Lett. B360 (1995) 13, hep-th/9508143.}.   

Although it might seem surprising that the Fradkin-Tseytlin term can
be written in manifestly SL(2,Z)-invariant form, one should remember
that the on-shell values of $S$ and $\tilde S$ spontaneously break
this SL(2,Z)-invariance. So the spacetime equations of motion coming
from the $\beta$-functions of the sigma model are not expected
to be SL(2,Z) invariant. 

Note that the terms $SF -\tilde S \tilde F$ and $( Sb'_{MN} -
\tilde S b_{MN}^{(0)}) \p_z 
Y^M \p_{\bar z} Y^N $ do not couple to the worldsheet metric components
$h_{zz}$ and $h_{\bar z \bar z}$, so $S$ and $\tilde S$
do not appear in the Virasoro generators. It seems reasonable to assume 
that they are absent also from the super-Virasoro generators, which
would imply that $S$ and 
$\tilde S$ are inert under worldsheet superconformal transformations. 

A curious feature of \signew\ is that 
it depends on two types of worldsheet gauge fields, one type coming
from D=2 N=(2,2) worldsheet supergravity and the other type coming from
the action of \town. It would be very interesting to find a 
relation between these two types of gauge fields.

\vskip 20pt

{\bf Acknowledgements:} I would like to thank Sergio Ferrara,
Warren Siegel and Paul Townsend for useful discussions and
CNPq grant 300256/94-9 for partial financial support.

\listrefs
\end